\documentclass[conference]{IEEEtran}
\IEEEoverridecommandlockouts
\usepackage{placeins}
\usepackage{amsmath,amssymb,amsfonts}
\usepackage{graphicx}
\usepackage{textcomp}
\usepackage{xcolor}
\usepackage{cite}
\usepackage{bm}
\usepackage{booktabs} 
\usepackage[skip=1pt]{caption}
\usepackage{pifont}

\usepackage{enumerate}
\usepackage{enumitem}
\usepackage[bookmarks=false]{hyperref}
\usepackage[
  paper=letterpaper,
  top=0.73in,
  bottom=1.01in,
  left=0.64in,
  right=0.64in,
  columnsep=0.25in
]{geometry}

\usepackage[capitalize]{cleveref}
\usepackage{mdframed}
\usepackage{framed} 
\newcommand{\rev}[1]{#1}

\begin{document}

\title{A Joint LLM and Optimization Agent Solution for Drone Scheduling in Victim Search Missions\vspace{-0.9cm}}

%\author{\IEEEauthorblockN{Anonymous Authors}}
\IEEEaftertitletext{\vspace{-1.0cm}}

\author{
    \IEEEauthorblockN{
    Yinghao Zhang\IEEEauthorrefmark{1}, 
    Yalin Liu\IEEEauthorrefmark{1}\P,
    Hong-Ning Dai\IEEEauthorrefmark{2},
    Yan Zhang\IEEEauthorrefmark{3},
    Qubeijian Wang\IEEEauthorrefmark{4}, 
    Bishenghui Tao\IEEEauthorrefmark{1}, and
    Yaohan Li\IEEEauthorrefmark{1}    
    }
    \IEEEauthorblockA{
        \IEEEauthorrefmark{1}%School of Science and Technology, 
        Hong Kong Metropolitan University, %\\
        \IEEEauthorrefmark{2}%Department of Computer Science, 
        Hong Kong Baptist University, Hong Kong, China \\
        \IEEEauthorrefmark{3}%School of Information and Communication Engineering, 
        University of Electronic Science and Technology of China, %\\
        \IEEEauthorrefmark{4}Northwestern Polytechnical University, China
    }
    \IEEEauthorblockA{Emails: \{zhangyi, ylliu, btao, yahli\}@hkmu.edu.hk, \{hndai, yanzhang\}@ieee.org, qubeijian.wang@nwpu.edu.cn}
    \thanks{\P~Corresponding author.}
    \thanks{This work was supported in part by the Research Grants Council of the Hong Kong Special Administrative Region, China, under Project UGC/FDS16/E15/24, in part by the Hong Kong Metropolitan University Research Grant under Project PFDS/2025/34, in part by the UGC Research Matching Grant Scheme under Project 2024/3003 and Gekko Lab, and in part by Hong Kong Metropolitan University Research Grant (No.2025/26 S\&T).}
}
% Yinghao Zhang, Yalin Liu, Hong-Ning Dai, Yan Zhang, Quenjian Wang, Bishenghui Tao, and Yaohan Li, "A Joint LLM and Optimization Agent Solution for Drone Scheduling in Victim Search Missions (under review)", The 22nd International Conference on Mobility, Sensing and Networking (MSN 2026), December 18-20, 2026, Ningbo, China.

\maketitle

\begin{abstract}
Victim search in wilderness areas is time-critical. Drones enable rapid area search, but scheduling them is challenging because victims may continue moving, detection results are uncertain, and effective decisions must combine structured mission information with unstructured contextual clues and domain knowledge. To address this, we propose an agent-aided drone scheduling system. The system employs an agent at the ground control station (GCS) to coordinate multiple drones over successive search cycles. In each cycle, the agent applies a scheduling policy that combines a large language model (LLM) for reasoning about where to search with an optimizer for determining how to dispatch the drones under energy constraints. Simulation results show that the proposed system consistently completes the victim-search mission, whereas none of the six non-LLM baselines does so and direct LLM scheduling is considerably less reliable. The source code is publicly available at \url{https://github.com/1643204431/llm-uav-victim-search}.
\end{abstract}

\vspace{-0.2cm}
\section{Introduction}
\label{sec:intro}
\vspace{-0.1cm}
Victim search in wilderness areas is highly time-critical, as every hour of delay dramatically reduces the probability of finding survivors alive~\cite{ge2025multi}. Additionally, the search areas often contain complex terrain that is difficult for ground search teams to access and traverse. The emerging drones offer strong maneuverability and cost-effectiveness, and are therefore promising for rapid victim-search missions~\cite{lahmeri2025uav, grassi2026geometrical, yan2026high}. A realistic mission typically involves multiple victims dispersed over a wide area, so multiple drones are required to simultaneously search different regions~\cite{li2025multiangle}. To complete the mission in time (i.e., confirm that all victims are found), a ground control station (GCS) is necessary to make global drone scheduling and coordinate their search activities~\cite{sharma2026cooperative, 11107240}. 

Existing studies have focused on drone scheduling based on approaches such as coverage path planning~\cite{choset2000coverage,lu2022coverage}, greedy dispatching to unexplored cells~\cite{burgard2005coordinated,yamauchi1998frontier}, and probabilistic belief-map updating~\cite{perez2018ant,perez2024minimizing}. These methods generally rely on limited and structured inputs, e.g., grid-map information and the number of victims. However, practical missions also involve unstructured prior victim-search knowledge, such as contextual clues from distress calls and domain knowledge from rescuers. Incorporating such unstructured knowledge into drone scheduling is both useful and largely underexplored.

Advanced large language models (LLMs) offer a promising way to integrate structured map information and unstructured prior knowledge for drone scheduling~\cite{wu2025llm,qayyum2025llm}. For instance, consider a distress call reporting ``the father is with his seven-year-old child; they heard running water.'' An LLM may infer that the pair tends to move toward a nearby water source and eventually remain close to it. The LLM can also use the child's limited walking range to exclude distant destinations and narrow the plausible combinations of routes and destinations. Despite this reasoning capability, LLMs alone cannot reliably produce drone scheduling decisions under a dynamic system status and strict energy constraints during long-term search~\cite{tong2025wirelessagent, qayyum2025llm, tan2026post}. As a consequence, directly generating an LLM-enabled drone scheduling decision is insufficient.

\begin{figure}[t]
\centering
\includegraphics[width=8.5cm]{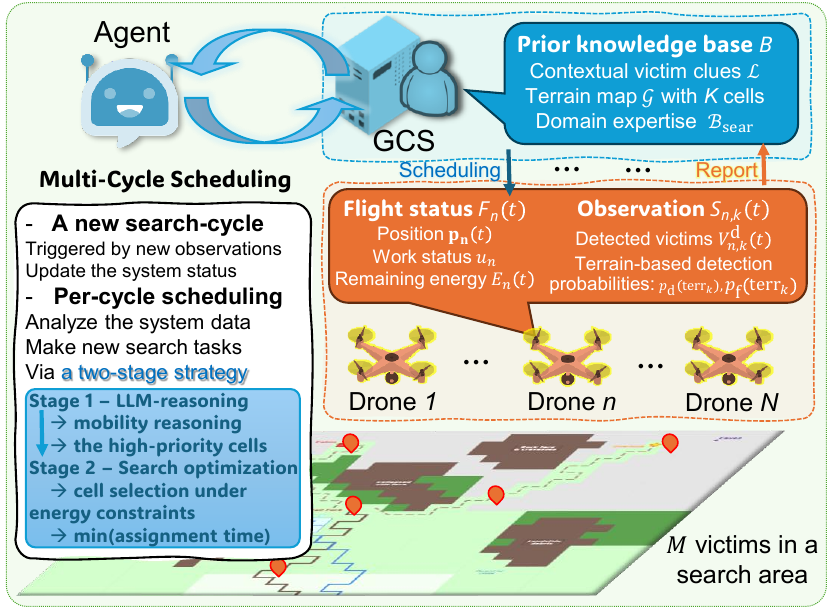}
\caption{\footnotesize The agent-aided drone scheduling system for victim search.}
\label{fig: system}
\vspace{-0.3cm}
\end{figure}

To bridge this gap, we aim to develop a joint LLM and optimization agent solution that utilizes an agent at the GCS to tightly couple LLM-based victim search reasoning with reliable optimization for drone scheduling. Designing such a joint global scheduling solution is non-trivial for three reasons. First, it is difficult to determine the effective search direction for drones because victim mobility is uncertain. Second, each drone is energy-constrained and cannot perform long-term search in a single flight. Third, each drone's observation changes over time. All these factors complicate the continuous global scheduling across multiple drones.

To address these issues, we design a comprehensive workflow for an agent-aided drone scheduling system, taking into account long-term system status tracking and continuous drone scheduling over multiple search cycles, as shown in \cref{fig: system}. In each cycle, the agent at the GCS applies a global scheduling policy to the current system state and prior knowledge to determine drone-to-cell assignments. The drones then fly to the assigned cells, perform search, and report new detection results. For efficient scheduling under dynamic status and drones' energy constraints, we develop a two-stage drone scheduling strategy that coordinates multiple energy-constrained drones over successive search cycles. In each cycle, the agent first (Stage 1) queries the LLM to reason over all available information and identify the most promising search cells. It then (Stage 2) invokes an optimizer to decide how to dispatch available drones to those selected cells with minimal time. Moreover, we evaluate the proposed system through simulations and show that it consistently completes the victim-search mission with substantially fewer search cycles and lower energy consumption than baseline methods.

\vspace{-0.1cm}
\section{System Workflow}
\label{sec:system}
\vspace{-0.1cm}
An agent-aided drone scheduling system is designed to execute victim search missions as shown in \cref{fig: system}. Given a coming mission that aims to search for victims in a target area, multiple drones are dispatched from a GCS to different regions of the area. Some prior clues of victims are known in the GCS. Each drone is equipped with an onboard sensing system to detect victims and continuously communicates with the GCS to report its observations, including search results and flight status. Considering the uncertain locations of victims and dynamic system status, the agent needs to make a global scheduling policy that can continuously analyze the global system status, infer the regions most likely to contain victims, and efficiently schedule drones to complete the mission. The detailed system workflow includes four parts: i) mission knowledge preparation, ii) drone configurations, iii) agent-aided global scheduling, and iv) two-stage scheduling strategy. 

\vspace{-0.2cm}
\subsection{Mission Knowledge Preparation}
\vspace{-0.1cm}
Let's consider a coming victim search mission where $M$ missing victims need to be found from a search area. The agent possesses some prior knowledge for this mission, stored as a database $\mathsf{B}$ represented by a three-element tuple as follows.
\vspace{-0.55cm}

\begin{align*}
    &\mathsf{B}= \left\langle\mathcal{L}, \mathcal{G}, \mathcal{B}_{\mathrm{sear}}\right\rangle,\text{where}\\
    &\mathcal{G} = \{\mathsf{G}_k|k\in\mathcal{K}\},\\
    &k\in\mathcal{K},\mathsf{G}_{k}=\left\langle \mathbf{p}_{k},\mathrm{terr}_{k} \right\rangle.
\end{align*}
\vspace{-0.55cm}

\noindent 
Herein, $\mathcal{L}$ denotes the collection of unstructured contextual clues of $M$ missing victims, which are available before the search. These clues may concern one or multiple victims and include victim profiles, group relationships, intended routes, and last-seen locations and times, e.g., ``{\footnotesize\ttfamily the child is 7 years old}''. The term $\mathcal{G}$ denotes the set of $K$ regular cells' terrain information for the search area. Terrain information of cell $k$ is a two-element tuple $\mathsf{G}_{k}=\left\langle \mathbf{p}_{k},\mathrm{terr}_{k} \right\rangle$, where $\mathbf{p}_{k}$ is the three-dimensional (3D) coordinate of its center and $\mathrm{terr}_{k}$ is its terrain type.
The term $\mathcal{K}=\{1,2,...,K\}$ is the index set of all search cells. The term $\mathcal{B}_{\mathrm{sear}}$ is the knowledge base that stores specialized domain information, including victim behavior profiles and terrain-mobility mappings that describe how terrain types affect movement speed and direction.
% rich cues about target location and mobility, but cannot be directly encoded into a conventional optimizer. Here, the agent has access to two such sources, both fixed throughout the mission.

\vspace{-0.2cm}
\subsection{Drone Configurations}
\vspace{-0.1cm}

\label{sec:uav}
Given $N$ drones in the GCS to be dispatched. Let $\mathcal{N}=\{1,2,...,N\}$ be the index sets of the drones. All drones adopt the same hardware configuration, including flight setting, onboard sensing, and the search task execution as follows. 

\subsubsection{Flight Setting} During the search, all drones are assumed to fly at a constant speed $v_c$ and the flight power $P_{\mathrm{f}}$, as their trajectories remain relatively stable at an altitude that ensures safe flight. At time $t$, the flight status of drone $n$ ($n\in\mathcal{N}$) is stored as a three-element tuple below:
    \vspace{-0.35cm}
    
    \begin{equation*}
        \mathsf{F}_{n}(t)=\left\langle \mathbf{p}_{n}(t),u_{n}(t),E_{n}(t) \right\rangle,
    \end{equation*}
    \vspace{-0.55cm}

    \noindent 
    where $\mathbf{p}_{n}(t)$ is its position, $u_{n}(t)\in \{0,1,...,K\}$ is the index of its scheduled search cell, and $E_{n}(t)$ is its remaining energy. The drone is active when $u_{n}(t)\neq 0$ and idle when $u_{n}(t)=0$, meaning that it has finished its scheduled task. Then, the estimated flight time required for drone $n$ to reach cell $k$ at time $t$ is $t_{n,k}=\|\mathbf{p}_{n}(t)-\mathbf{p}_{k}\|/v_c$.
    
\subsubsection{Onboard sensing} The onboard sensing system applies a vision-based detection model pretrained for the traversable terrain types considered in the search area. The model outputs the number of victims detected in each cell, and its power consumption is fixed at $P_{\mathrm{d}}$. Its detection performance is characterized by two terrain-dependent probabilities, $p_{\mathrm{d}}(\mathrm{terr}_{k})$ and $p_{\mathrm{f}}(\mathrm{terr}_{k})$, denoting the probability of correctly detecting a victim present in the cell and the probability of falsely reporting a victim when none is present, respectively. Each positive detector output is verified by the agent. 

\subsubsection{Search Task Execution}
\label{sec:task}
During the search mission, $N$ drones continuously execute the search tasks scheduled by the agent. Given a scheduled search task that the $n$-th drone ($n\in\mathcal{N}$) searches the $k$-th cell ($k\in\mathcal{K}$), it is executed via two steps: i), the $n$-th drone flies to the center of the $k$-th cell; and ii) the drone hovers and detects the presence of victims upon arrival. The detection outcome is stored as a three-element tuple:
\vspace{-0.15cm}
\begin{equation*}
\mathsf{S}_{n,k}(t)=\left\langle V_{n,k}^{\mathrm{d}}(t), p_{\mathrm{d}}(\mathrm{terr}_{k}), p_{\mathrm{f}}(\mathrm{terr}_{k}) \right\rangle,
\end{equation*}%
\vspace{-0.55cm}

\noindent 
where $V_{n,k}^{\mathrm{d}}(t)\in \{0,1,2,...,M\}$ records the number of victims detected by the onboard model. Once victims are found, the drone instructs them to remain in place and reports their locations to the agent.

\vspace{-0.3cm}
\subsection{Agent-aided Global Scheduling}
\label{sec:cycle}
\vspace{-0.1cm}
To find all $M$ victims in the target mission, the agent continuously performs global scheduling by analyzing the global system status and dispatching drones to new cells. As the system status changes, the global scheduling is performed across multiple analyze-dispatch cycles. Each cycle determines multiple search tasks that pair idle drones and search cells. 

To ensure secure drone recall in searching a specific cell (say the $k$-th cell), the agent needs to find an idle drone (say the $n$-th drone) that satisfies a sufficient minimum value for the remaining energy $E_{n}(t)$, called a \textit{Effective Energy Reserving Condition (EERC)} defined as follows.
\vspace{-0.5cm}  

\begin{align*}
    \mathsf{EERC}: &E_{n}(t)\geq E_{n,k}^{\mathrm{f}}(t)+E_{k}^{\mathrm{d}} +E_{n,k}^{\mathrm{re}},\text{ where}\\
    &E_{n,k}^{\mathrm{f}}(t) = P_{\mathrm{f}} \|\mathbf{p}_{n}(t) - \mathbf{p}_{k}\|/v_c,
    E_{k}^{\mathrm{d}} = P_{\mathrm{d}}\hat{T}_{k}^{\mathrm{d}}(t),\\
    &E_{n,k}^{\mathrm{re}} = P_{\mathrm{f}}\|\mathbf{p}_k - \mathbf{p}_{\text{base}}\|/v_c+ E_{\text{buf}},
\end{align*}%
where $E_{n,k}^{\mathrm{f}}$ is the energy consumed during the flight to the cell, $E_{k}^{\mathrm{d}}$ is the energy consumed while searching the cell, $E_{n,k}^{\mathrm{re}}$ is the energy reserved for safely returning to the base point at $\mathbf{p}_{\text{base}}$, $\hat{T}_{k}^{\mathrm{d}}(t)$ is the expected cell-search time, determined by the cell area and terrain complexity, and $E_{\text{buf}}$ is a fixed safety margin for all drones. 

Accordingly, the agent runs the per-cycle scheduling below. 
\begin{itemize}[leftmargin=*,align=left]
    \item 
    First, the cycle is triggered when the agent receives at least one new observation (i.e., the drone's detection results) and a certain number (denoted by $B$) of drones become idle. Then the agent %receives at least one new observation and 
    updates the global system status data $\mathsf{O}(t)$, which is stored as a four-element tuple below:
    \vspace{-0.1cm}
    \begin{align*}
        & \mathsf{O}(t) 
        =\left\langle \hat{V}_{\mathrm{global}}(t),\, \hat{\mathcal{S}}(t),\, \hat{\mathcal{A}}(t), \, \hat{\mathcal{F}}(t)\right\rangle,\text{ where}\\
        & \hat{V}_{\mathrm{global}}(t)\in\{0,1,2,...,M\},\hat{\mathcal{F}}(t)=\{\mathsf{F}_{n}(t)|n\in\mathcal{N}\},\\
        & \hat{\mathcal{S}}(t)=\{\mathsf{S}_{n,k}(\tau)a_{n,k}(\tau)|\tau\in [0,t],n\in\mathcal{N},k\in\mathcal{K}\},\\
        & \hat{\mathcal{A}}(t)=\{a_{n,k}(t)\in\{0,1\}|n\in\mathcal{N},k\in\mathcal{K}\},
    \end{align*}
    
    \vspace{-0.1cm}
    
    \noindent 
    \rev{where $\hat{V}_{\mathrm{global}}$ records the number of confirmed victims, $\hat{\mathcal{S}}$ records all historical detection results, $\hat{\mathcal{A}}$ is the set of ongoing search tasks, and $\hat{\mathcal{F}}(t)$ collects the flight status of all drones. In $\hat{\mathcal{A}}$, $a_{n,k}$ is a binary indicator that equals $1$ when the $k$-th cell is being searched by the $n$-th drone, corresponding to the search status $u_{n}(t)=k$ in $\mathsf{F}_{n}$. 
    }
    
    \item 
    The agent then analyzes the current system data $\mathsf{O}(t)$ and the victim search prior knowledge $\mathsf{B}$, and makes new search tasks via a per-cycle scheduling policy defined by
    \begin{align*}
        &\pi(t): (\mathsf{O}(t), \mathsf{B}) \rightarrow \tilde{\mathcal{A}}(t),
        \\
        &\tilde{\mathcal{A}}(t) = \{\tilde{a}_{n,k}(t)=1\mid k\in\tilde{\mathcal{K}}_{\mathsf{sear}}(t),n\in\tilde{\mathcal{N}}_{\mathsf{disp}}(t)\},
    \end{align*}%
    where $\tilde{\mathcal{A}}(t)$ is a set of scheduled search tasks with $\tilde{a}_{n,k}(t)=1$ indicating that the $n$-th drone is scheduled to search the $k$-th cell. The term $\tilde{\mathcal{K}}_{\mathsf{sear}}(t)$ is the selected search cells and $\tilde{\mathcal{N}}_{\mathsf{disp}}(t)$ is the $B$ drones to be dispatched. The dispatched drone set $\tilde{\mathcal{N}}_{\mathsf{disp}}(t)$ consists of idle drones that must satisfy $\mathsf{EERC}$ for at least one candidate cell in $\tilde{\mathcal{K}}_{\mathsf{sear}}(t)$. Idle drones that do not satisfy $\mathsf{EERC}$ are recalled to the base point. Accordingly, the dispatched drones fly to the search cells, execute victim detection, and report their observations. 
\end{itemize}
Overall, the global scheduling in the agent evolves across multiple cycles as new observations arrive. The search is completed when all $M$ victims have been confirmed found; otherwise, it terminates when no feasible drone dispatch remains or a prescribed cycle cap is reached. %The global scheduling policy is determined by minimizing the mission completion time, as given in~\cref{subsec: problem}.

\vspace{-0.2cm}
\subsection{Two-Stage Scheduling Strategy}
\label{subsec: problem}
\vspace{-0.1cm}
To run a victim search mission efficiently, the multi-cycle global scheduling (mentioned in \cref{sec:cycle}) needs to be completed (i.e., confirm all victims found) in a minimum time. The terms $T_0$ and $T_{\mathrm{comp}}$ denote the start time\footnote{The value of $T_0$ is the response delay from the received request for the mission to the start of the first search cycle, which is generally determined by the system's running latency during the warm-up.} and the completion time of the search mission. Suppose that all victims are confirmed found in $C$ cycles, the mission completion time $T_{\mathrm{comp}}$ can be calculated by
$T_{\mathrm{comp}}=T_0+\sum_{c=1}^{C}\left(T_{\mathrm{agent}}^{(c)}+T_{\mathrm{idle}}^{(c)}\right)$,
where %$T_0$ is the start time of the first search cycle; 
$T_{\mathrm{agent}}^{(c)}$ and $T_{\mathrm{idle}}^{(c)}$ are the agent computation time and the time from the completion of the agent's decision to the beginning of the subsequent cycle, respectively. Let $\mathcal{T}_{\mathrm{dec}}=\{t_c\mid c=1,\ldots,C\}\subseteq[T_0,T_{\mathrm{comp}}]$ denote the set of decision times. Let $\Pi=\{\pi(\tau)|\tau\in \mathcal{T}_{\mathrm{dec}}\}$ denote the set of per-cycle drone-scheduling policies used to complete the mission.
% Therefore, a multi-cycle drone scheduling problem $\mathcal{P}_{\mathrm{1}}$ is formulated by minimizing $T_{\mathrm{comp}}$ while satisfying EERC.
%\vspace{-0.4cm}  
Then, an efficient $\Pi$ needs to be found by minimizing $T_{\mathrm{comp}}$ as follows. 
\begin{subequations}
\label{eq:problem}
\begin{align}
    &%\mathcal{P}_{\mathrm{1}}: 
    \min_{\Pi} \quad T_{\mathrm{comp}}
    % = \mathbb{E}_{\Pi}[\max_{m\in\mathcal{M}} \tilde{T}_m]
    \label{eq:objective} \\
    &\text{s.t.} \quad
    \forall \tau\in \mathcal{T}_{\mathrm{dec}},\pi(\tau): (\mathsf{O}(\tau), \mathsf{B}) \rightarrow \tilde{\mathcal{A}}(\tau),\notag \\
    & \tilde{\mathcal{K}}_{\mathsf{sear}}(\tau) \subseteq \mathcal{K}|_{\sum_{n} a_{n,k}(\tau)=0},
    \tilde{\mathcal{N}}_{\mathsf{disp}}(\tau) \subseteq \mathcal{N}|_{\sum_{k} a_{n,k}(\tau)=0},\label{eq:occupied}\\%a_{n,k}(t)\in\mathsf{O}(t)
    %& \tilde{\mathcal{N}}_{\mathsf{disp}}(\tau) \subseteq  \{n| \sum_{k} a_{n,k}(\tau)=0\},\\ %a_{n,k}(t)\in\mathsf{O}(t)
    & \sum_{k\in\tilde{\mathcal{K}}_{\mathsf{sear}}(\tau)} \tilde{a}_{n,k}(\tau) \leq 1, 
    \sum_{n\in\tilde{\mathcal{N}}_{\mathsf{disp}}(\tau)} \tilde{a}_{n,k}(\tau)\leq 1, \label{eq:ontoone} \\
    & \forall\, \tilde{a}_{n,k}(\tau)\in\tilde{\mathcal{A}}(\tau),
    \mathsf{EERC}|_{\tilde{a}_{n,k}(\tau)}.\label{eq:energy}
    \end{align}
\end{subequations}

\noindent 
Herein, constraint~\eqref{eq:occupied} prevents idle drones from being assigned to cells already occupied by active drones, \eqref{eq:ontoone} enforces one-to-one assignments, and~\eqref{eq:energy} ensures sufficient energy for returning to the base with the required safety reserve after completing the search. It can be observed that $T_{\mathrm{comp}}$ is affected by both the number of search cycles and the duration of each cycle. The number of search cycles primarily depends on which cells are selected for search in each cycle. For example, if the selected cells are more likely to contain unfound victims, the mission can be completed in fewer search cycles. The duration of each cycle is affected by how the drones are assigned to the selected cells. 

Motivated by the above, the global scheduling policies $\Pi$ can be found in a two-stage scheduling strategy in each cycle:

\begin{mdframed}[innerleftmargin=0pt, leftmargin=0pt]
\small
\begin{itemize}[leftmargin=*,align=left]
\item[] {\scriptsize \#\texttt{A two-stage scheduling strategy}\vspace{-0.2cm}}
\rule{\linewidth}{0.4pt}
    \item[] \textit{Stage 1 - Cell prioritization:} In each cycle at time $t$, given the current system state $\mathsf{O}(t)$ and prior knowledge $\mathsf{B}$, the agent identifies the probability of containing unfound victims in cells and assigns them a priority vector $\mathbf{w}(t)\in[0,1]^K$, where $w_k(t)$ represents the relative search priority of cell $k$. %This will assist in determining the effective number of search cycles through the quality of the generated cell priorities.
    \item[] \textit{Stage 2 - Efficient assignment:}
    Given $\mathbf{w}(t)$ and the scheduling constraints in
    \eqref{eq:occupied}--\eqref{eq:energy}, the agent assigns drones in the current dispatch batch to top-$B$ high-priority reachable cells while minimizing the worst-case outbound travel time. This affects the search tasks of each cycle through the drone-to-cell assignment. 
\end{itemize}
\end{mdframed}

%\vspace{-0.1cm}
\section{Joint LLM and Optimization Implementation}
\label{sec:agent}
%\vspace{-0.1cm}
In order to implement the two-stage scheduling strategy, the agent needs to analyze the per-cycle system status and generate the efficient per-cycle policy. Thus, we present a joint LLM and Optimization solution running two stages. In Stage 1, the agent queries an LLM to reason about the priority weight vector. In Stage 2, the agent invokes a drone-to-cell optimizer to minimize the worst-case travel time to high-priority cells.

%Based on the above objective, we develop a two-stage per-cycle policy. To reduce cycles needed to find all victims, Stage~1 uses an LLM to combine structured map and system-state information with unstructured contextual clues and domain knowledge, assigning higher priorities to cells more likely to contain unfound victims. To reduce the controllable portion of the per-cycle duration, Stage~2 uses an optimizer to select energy-feasible high-priority cells and assign idle drones while minimizing the worst-case outbound travel time.
% -------------------------------------------------
\vspace{-0.1cm}
\subsection{Stage 1: LLM Reasoning}
\label{sec:cot}
The agent in this stage reasons about the priority weight vector $\mathbf{w}(t)$ based on the contextual clues $\mathcal{L}$ and the domain expertise $\mathcal{B}_{\mathrm{sear}}$. 
%Because the retrieval is performed once and reused across all search cycles, it adds minimal per-cycle latency. 
%Directly mapping system status and prior knowledge to a numerical priority vector in one step is difficult, as the LLM must simultaneously interpret clues, reason about victim behavior, and assess every cell. 
Particularly, the agent queries an LLM to perform the knowledge preparation and a two-step reasoning: 
% first building a qualitative understanding of each target's mobility and then translating that understanding into quantitative cell scores.
% that mirrors expert victim search reasoning
% Below is the design of the two steps.

\begin{itemize}[leftmargin=*,align=left]
\item \textit{Relevant knowledge preparation.} To provide relevant domain expertise to the LLM, the agent first uses the contextual clues $\mathcal{L}$ as queries to retrieve a mission-relevant subset $\mathcal{R}$ from the specialized domain knowledge $\mathcal{B}_{\mathrm{sear}}$. The agent then combines the retrieved set $\mathcal{R}$, terrain map $\mathcal{G}$, and clue set $\mathcal{L}$ as the relevant prior knowledge $\mathsf{B}|_{\mathcal{R}}$.
\item \textit{Step~1: Mobility Pattern Reasoning.}
\rev{The LLM reasons about each unfound victim's mobility pattern and likely current location based on the system status $\mathsf{O}(t)$ and relevant prior knowledge $\mathsf{B}|_{\mathcal{R}}$, producing a natural-language analysis.}
\item \textit{Step~2: Cell Prioritization.}
The LLM maps the analysis from Step~1 onto the search area to produce the priority vector that identifies the search cells in which the victims are most likely to appear and quantifies their priorities using a weight vector $\mathbf{w}(t) \in [0,1]^K$. To save token, only the top-$K'$ ($K' \ll K$) cells will be reported by LLM.
\end{itemize}

%\subsubsection{Prompt Design}
A prompt is designed to run the above reasoning, consisting of a system message and a user message as follows.

\begin{mdframed}[innerleftmargin=0pt, leftmargin=0pt]
\vspace{-0.1cm}
{\footnotesize
\begin{itemize}[leftmargin=*,align=left,nosep]
\item[] {\scriptsize \#\texttt{LLM Reasoning Prompt Design}\vspace{-0.2cm}}
\rule{\linewidth}{0.4pt}
\item[] \textbf{System Message} {\scriptsize \#\texttt{assembled once at mission start} }
\begin{itemize}[leftmargin=*,nosep]
\item \textit{Role and task.} \rev{You are the victim search agent deployed at the GCS. At search cycle $t$, solve the cell-prioritization problem by estimating how likely each search cell is to contain unfound victims. Represent the result using $\mathbf{w}(t)\in[0,1]^K$, where $w_k$ is the priority of cell $k$. Do not determine drone-to-cell assignments; they are computed by the optimizer in Stage~2.}
\item \textit{Relevant prior knowledge.} \rev{Use the terrain map $\mathcal{G}$, contextual clues $\mathcal{L}$, and retrieved domain knowledge $\mathcal{R}$ provided below as $\mathsf{B}|_{\mathcal{R}}$.}
\item \textit{Step 1: Mobility Pattern Reasoning.} \rev{For each unfound victim, infer plausible mobility patterns and current locations using the contextual clues, elapsed time, terrain and trail structure, landmarks, retrieved walking-speed and stamina ranges, and accumulated detection history. When the clues indicate that multiple victims travel together, reason about them jointly and bound their movement using the least-mobile member. Use physical reachability to exclude locations that the victims could not have reached.}
\item \textit{Step 2: Cell Prioritization.} \rev{Map the mobility analysis from Step~1 onto the cell set $\mathcal{K}$. Assign larger $w_k$ to cells that are more consistent with the inferred routes, destinations, rest locations, and current observations. Revise previously searched cells using their detection records, while distinguishing sensor observations from confirmed victim reports.}
\item \textit{Behavior principles.} \rev{Impassable cells cannot contain victims and must receive zero priority. Terrain and impassable regions may nevertheless constrain the victims' feasible movement routes. Confirmed victims are excluded from subsequent reasoning. A negative detection reduces the priority of the corresponding cell but does not by itself prove that the cell is empty.}
\item \textit{Output format.} \rev{Provide the mobility analysis from Step~1, followed by exactly one JSON object containing the top-$K'$ entries of $\mathbf{w}(t)$: \texttt{\{"priorities": \{"row,col": w\_k, ...\}\}}. Each coordinate must refer to a valid, passable cell, and every reported score must lie in $[0,1]$. Cells not listed in the JSON object are assigned zero priority.}
\end{itemize}
\item[] \textbf{User Message} {\scriptsize \#\texttt{rebuilt at each search cycle}}
\item[] \rev{Current cycle time: $t$. Remaining unfound victims: $M\!-\!\hat{V}_{\mathrm{global}}(t)$. Confirmed victims: \textit{their locations and confirmation times}. Detection history: $\hat{\mathcal{S}}(t)$.}
\end{itemize}
}
\end{mdframed}

Once at mission start, the system message is constructed with four parts: 
(i)~a role definition with two-step reasoning instructions; 
(ii)~the behavior principles of victims embedded as text. \rev{With only general knowledge in the agent, the above reasoning cannot cater to specific victims' physical behaviors (e.g., victims cannot appear in physically unreachable cells).}; (iii)~the relevant prior knowledge $\mathsf{B}|_{\mathcal{R}}$, including $\mathcal{G}$, $\mathcal{L}$, and $\mathcal{R}$, embedded as plain text; and (iv)~a structured output format that specifies the JSON schema for the LLM to output $\mathbf{w}(t)$ for the top-$K'$ cells.
The user message is rebuilt in each search cycle using four fields of cycle-varying information: (i)~the current cycle time $t$, (ii)~the remaining unfound victim count $M - \hat{V}_{\mathrm{global}}(t)$, (iii)~the locations and confirmation times of confirmed victims, and (iv)~the detection history $\hat{\mathcal{S}}(t)$.

Within a search cycle, if the LLM response is unusable (e.g., it reaches the output-token limit or cannot be parsed into a valid priority vector), the agent queries the LLM again using the same prompt. If the second response remains unusable, the agent applies a fallback that assigns the same priority to every passable cell for that search cycle.

\vspace{-0.2cm}
\subsection{Stage 2: Drone-to-Cell Optimization}
\label{sec:step3}
\vspace{-0.1cm}
After the above reasoning, the agent obtains the top-$K'$ cells and invokes an optimizer to determine drone-to-cell assignments. It first selects a small candidate cell set from $\mathbf{w}(t)$ and then solves a bottleneck assignment over that set by minimizing the worst-case travel time. 
\vspace{-0.1cm}
\begin{itemize}[leftmargin=*,align=left]
\item \textit{Cell selection.} \rev{For each drone $n$ in the current dispatch batch $\tilde{\mathcal{N}}_{\mathsf{disp}}(t)$, its reachable cell set $\mathcal{R}_n(t)$ contains cells that satisfy two conditions: each cell is not currently being searched by another drone, and the drone has sufficient energy to fly to the cell, complete the search, and return to the base, i.e.,}
\vspace{-0.6cm}

\begin{align*}
\mathcal{R}_n(t) = \left\{k\in\mathcal{K}\;\middle|\;
\begin{array}{l}
\textstyle\sum_{n'} a_{n',k}(t)=0,\\[-0.1mm]
\mathsf{EERC}\text{ holds for }(n,k)
\end{array}\right\}.
\end{align*}%
\vspace{-0.3cm}

\noindent 
The candidate cell pool $\mathcal{R}^*(t)=
\bigcap_{n\in\tilde{\mathcal{N}}_{\mathsf{disp}}(t)}
\mathcal{R}_n(t)$ retains only the cells reachable by every drone in the current dispatch batch. The optimizer selects the top-$B$ cells from $\mathcal{R}^*(t)$ according to $w_k(t)$ to form a candidate set $\mathcal{C}(t)$. %If $|\mathcal{R}^*(t)| < B$, then $\mathcal{C}(t) = \mathcal{R}^*(t)$.

\item \textit{Assignment Time Minimization.} %A bottleneck assignment is used to determine the dispatched drones with the minimum worst-case travel time. 
\rev{The optimizer constructs an efficient assignment problem $\tilde{\mathcal{P}}_{\mathrm{1}.2}$ to find a maximum-cardinality assignment $\tilde{\mathcal{A}}(t)$ while minimizing the worst-case travel time to the top-$B$ cells. The problem is given by}
%$\mathcal{P}_{\mathrm{1}.2}$

\vspace{-0.65cm}
\begin{subequations}
\begin{align}
    %\tilde{\mathcal{P}}_{\mathrm{1}.2}: 
    \min_{\{\tilde{a}_{n,k}\}} & \quad 
\max_{n \in \tilde{\mathcal{N}}_{\mathsf{disp}}(t),\, 
      k \in \mathcal{C}(t)} 
\; t_{n,k} \cdot \tilde{a}_{n,k}(t)
    % \min_{\tilde{\mathcal{A}}(t)} \quad
    % & \max_{n,k:\, \tilde{a}_{n,k}(t)=1} t_{n,k}
    \label{eq:bottleneck_obj}\\
    \text{s.t.} \quad 
    & \sum_{n,k} \tilde{a}_{n,k}(t) = \min\!\big(B,\, |\mathcal{C}(t)|\big),
    \label{eq:bottleneck_card}\\
    & \sum_{k\in\mathcal{C}(t)} \tilde{a}_{n,k}(t) \leq 1,
    %\quad \forall\, n:\, u_n(t)=0,
    %\label{eq:bottleneck_uav}\\
    %& 
    \sum_{n\in\tilde{\mathcal{N}}_{\mathsf{disp}}(t)}
\tilde{a}_{n,k}(t) \leq 1. %,
    %\quad \forall\, k\in\mathcal{C}(t).
    \label{eq:bottleneck_cell}
\end{align}
\end{subequations}
\vspace{-0.4cm}

\noindent 
%where $ t_{n,k}$ a travel-time matrix with entries. 
\rev{$\tilde{\mathcal{P}}_{\mathrm{1}.2}$ is solved by binary search over the sorted set of finite travel times. Specifically, given a travel-time threshold $\sigma$, a bipartite graph retaining only edges with $t_{n,k} \leq \sigma$ is constructed, the maximum matching is computed, and the smallest $\sigma$ that yields the cardinality required by~\cref{eq:bottleneck_card} is selected. Finally, the resulting $\tilde{\mathcal{A}}(t)$ is the output of the per-cycle policy $\pi(t)$. Drones not assigned in $\tilde{\mathcal{A}}(t)$ remain idle and participate in the next search cycle.}
\end{itemize}

% \textit{Performance analysis:} The above two-stage scheduling strategy affects the mission performance as follows. First, in the cell-prioritization stage, an upper boundary for the expected completion cycle exists, and it is even tighter if the priority vector is set with a larger lower bound $\underline{\rho}$ for a given onboard detection model\footnote{The proof is provided in Appendix \ref{proof_the1}.}. Meanwhile, given the cells selected for search in the assignment stage, different drone-to-cell assignments result in different outbound travel times. We characterize this part of the scheduling performance by the worst-case outbound travel time $\max_{n,k:\tilde a_{n,k}(t_c)=1}t_{n,k}$. A smaller value reduces the maximum travel delay before the assigned search cells are reached. Overall, it can be seen that the two stages affect the mission performance through two different quantities: $\rho_{m,c}(\mathbf w(t_c))$ characterizes the cell-prioritization performance related to the number of search cycles, while the worst-case outbound travel time characterizes the assignment efficiency related to the per-cycle travel delay.

\textit{Performance analysis:} The above scheduling strategy affects the mission performance as follows. First, in the cell-prioritization stage, an upper bound for the expected completion cycle exists, and this upper bound decreases when the top-$B$ high-priority reachable cells are more likely to contain unfound victims. The derivation is provided in Appendix~\ref{proof_the1}. Meanwhile, given the selected cells in the assignment stage, %different drone-to-cell assignments result in different outbound travel times. We characterize this part of 
the scheduling performance can be characterized by the worst-case outbound travel time $\max_{n,k:\tilde a_{n,k}(t_c)=1}t_{n,k}$. A smaller value reduces the maximum travel delay before the assigned search cells are reached. 

% Overall, it can be seen that the two stages affect the mission performance through two different quantities: $\rho_{m,c}(\mathbf w(t_c))$ characterizes the cell-prioritization performance related to the number of search cycles, while the worst-case outbound travel time characterizes the assignment efficiency related to the per-cycle travel delay.

\vspace{-0.3cm} 
\section{Numerical Results} \label{sec:experiments}

\vspace{-0.2cm}
\subsection{Simulation Setup}
\label{sec:setup}
\vspace{-0.1cm}

\subsubsection{Map and Contextual Clues}
We simulate a $7\,\text{km}\times7\,\text{km}$ wilderness search area discretized into $28\times28$ equal-sized cells. %, as shown in \cref{fig:case}(a). 
The map $\mathcal{G}$ is available to the agent before the search, containing eight terrain types and specifying the locations of GCS/Entry (the southern trailhead), Trail Fork, Overlook, Caves, Pond, and Campsite. 
%\subsubsection{Contextual Clues}
%In the simulation, 
Given a victim search mission, the exact clue text provided to the agent is as follows.

\begin{mdframed}[
linewidth=0.4pt,
leftmargin=0pt,
rightmargin=0pt,
innerleftmargin=4pt,
innerrightmargin=4pt,
innertopmargin=3pt,
innerbottommargin=3pt,
skipabove=3pt,
skipbelow=3pt
]
{\scriptsize\ttfamily\raggedright
A hiking group of 5 people departed from the southern trailhead this morning at 08:30, heading for the overlook in the northeast of the region as their shared destination. They were expected back by 5 PM but have not returned and are now unreachable by phone. Group composition: Family A (a parent in their 40s with a 7-year-old child; the young child limits how far the family can travel), Family B (a married couple in their early 30s, moderately fit), and Youth C (a 22-year-old experienced solo hiker with strong stamina). Last visual sighting: a passing hiker saw all five at the trail fork around 10:00 AM; no sightings since.
\par}
\end{mdframed}

\subsubsection{Ground-truth mobility} Using the same terrain map, the simulator generates the victims' subsequent trajectories as the simulation ground truth. These trajectories, including the subgroup-specific destinations and mobility parameters, are not provided to the agent. After reaching Trail Fork, \textit{Family~A} heads toward Pond, \textit{Family~B} heads toward Campsite, and \textit{Youth~C} continues toward Overlook. The three subgroups follow terrain-weighted shortest paths that favor marked trails, at constant speeds of $0.5$, $0.8$, and $1.0\,\text{m/s}$, respectively. Upon arrival, they wander within radii of $300$, $400$, and $500\,\text{m}$ at reduced speeds of $0.2$, $0.3$, and $0.4\,\text{m/s}$, respectively. The wandering direction varies with the random seed.

\subsubsection{Drone Configuration}
% \rev{We configure the fleet size, energy budget, onboard detection model, and terrain-dependent search capability of the drones.}
The default number of drones is $N=10$. Each drone has initial energy $E_{\mathrm{init}}=500\,\text{kJ}$, safety reserve $E_{\mathrm{buf}}=30\,\text{kJ}$, cruise speed $v_c=15\,\text{m/s}$, flight power $P_{\mathrm{f}}=200\,\text{W}$, and search power $P_{\mathrm{d}}=150\,\text{W}$. The dispatch batch size is $B=10$ in each search cycle. The terrain-dependent detection parameters $(p_{\mathrm{d}},p_{\mathrm{f}})$ are set to $(0.95,0.02)$ for meadow, trail, creek, and pond, $(0.80,0.05)$ for sparse vegetation, $(0.55,0.10)$ for dense forest, and $(0.65,0.12)$ for rocky terrain. The cell-search times are $50$, $60$, $60$, $70$, $80$, $100$, and $120\,\text{s}$ for trail, meadow, pond, creek, sparse vegetation, rocky terrain, and dense forest, respectively.

\subsubsection{Agent Configuration}
% \rev{We configure the mission timeline, retrieved domain knowledge, and LLM backbones used by the agent.}
We take the last sighting at $10{:}00$\,AM as $t=0$. After a response delay of $T_0=3600\,\text{s}$, the drones are dispatched and the first search cycle begins at $t=T_0$ (i.e., $11{:}00$\,AM). The retrieved knowledge $\mathcal{R}$ contains category-specific on-trail/off-trail walking-speed and stamina ranges: $0.4$--$0.7/0.2$--$0.4\,\text{m/s}$ and $4$--$5\,\text{km}$ for children, $0.7$--$1.0/0.4$--$0.6\,\text{m/s}$ and $6$--$9\,\text{km}$ for average adults, and $0.9$--$1.3/0.6$--$0.9\,\text{m/s}$ and $15$--$30\,\text{km}$ for experienced hikers. It also includes water-seeking and physical-reachability heuristics. The default backbone is DeepSeek V4 Pro in non-thinking mode. DeepSeek V4 Flash, Qwen Flash, Qwen Plus, and Qwen3 Max are evaluated under the same non-thinking setting for comparison. We set $K'=50$, so the LLM reports the 50 highest-priority entries of $\mathbf{w}(t)$ in each search cycle; unlisted cells are assigned zero priority.

\begin{table*}[t]
\centering
\caption{Architecture and component ablation results. Bold marks the best result in columns that permit direct comparison.}
\label{tab:ablation}
\setlength{\tabcolsep}{2.0pt}
\resizebox{\textwidth}{!}{%
\scriptsize
\begin{tabular}{@{}lccccccccc@{}}
\toprule
\textbf{Method / Variant}
& $\mathrm{Found}$
& $\mathrm{Compl.}$
& $\mathrm{Cycles}$
& $T_{\mathrm{comp}}$ (s)
& $E/\mathrm{vic}$ (kJ)
& $\mathrm{Tok/run}$
& $T_{\mathrm{agent}}/\mathrm{cyc.}$ (s)
& Illegal
& $\mathrm{Fb.}$ \\
\midrule

{\scriptsize\texttt{Ours}} ({\scriptsize DS V4 Pro})
& $\mathbf{5.0\pm0.0}$
& $\mathbf{5/5}$
& $\mathbf{3.8\pm1.8}$
& $\mathbf{4661\pm457}$
& $\mathbf{323\pm128}$
& $\mathbf{32k\pm16k}$
& $22\pm1$
& $\mathbf{26.3\%}$
& $\mathbf{0}$ \\

{\scriptsize\texttt{LLM-Direct}}
& $4.0\pm1.2$
& $2/5$
& $28.0\pm11.0$
& $6223\pm1083$
& $1160\pm729$
& $273k\pm127k$
& $\mathbf{9\pm0}$
& $52.1\%$
& $6$ \\

{\scriptsize\texttt{w/o Knowledge Retrieval}}
& $4.6\pm0.9$
& $4/5$
& $9.6\pm12.5$
& $4654\pm264$
& $556\pm560$
& $95k\pm130k$
& $24\pm9$
& $38.8\%$
& $\mathbf{0}$ \\

{\scriptsize\texttt{w/o Mobility Reasoning}}
& $3.2\pm1.5$
& $1/5$
& $26.0\pm12.5$
& $4682\pm0$
& $1854\pm1654$
& $289k\pm162k$
& $17\pm7$
& $33.8\%$
& $5$ \\

{\scriptsize\texttt{w/o Bottleneck}}
& $\mathbf{5.0\pm0.0}$
& $\mathbf{5/5}$
& $\mathbf{3.8\pm1.8}$
& $4703\pm453$
& $327\pm136$
& $\mathbf{32k\pm16k}$
& $22\pm1$
& $\mathbf{26.3\%}$
& $\mathbf{0}$ \\

\bottomrule
\end{tabular}%
}
\end{table*}

\begin{table*}[t]
\vspace{-0.2cm}
\centering
\caption{Cross-backbone results for {\scriptsize\texttt{Ours}}.}
\label{tab:backbone}
\setlength{\tabcolsep}{2.05pt}
\tiny
\resizebox{\textwidth}{!}{%
\begin{tabular}{@{}lccccccccc@{}}
\toprule
\textbf{Backbone}
& $\mathrm{Found}$
& $\mathrm{Compl.}$
& $\mathrm{Cycles}$
& $T_{\mathrm{comp}}$ (s)
& $E/\mathrm{vic}$ (kJ)
& $\mathrm{Tok/run}$
& $T_{\mathrm{agent}}/\mathrm{cyc.}$ (s)
& Illegal
& $\mathrm{Fb.}$ \\
\midrule
DeepSeek V4 Pro
& $5.0\pm0.0$
& $5/5$
& $3.8\pm1.8$
& $4661\pm457$
& $323\pm128$
& $32k\pm16k$
& $22\pm1$
& $26.3\%$
& $0$ \\

DeepSeek V4 Flash
& $4.8\pm0.4$
& $4/5$
& $5.0\pm6.7$
& $4399\pm253$
& $406\pm428$
& $91k\pm131k$
& $54\pm21$
& $12.5\%$
& $1$ \\

Qwen Flash
& $1.6\pm0.9$
& $0/5$
& $40.0\pm0.0$
& ---
& $2201\pm65$
& $566k\pm9k$
& $19\pm2$
& $93.0\%$
& $0$ \\

Qwen Plus
& $5.0\pm0.0$
& $5/5$
& $5.6\pm3.2$
& $6426\pm1536$
& $470\pm242$
& $114k\pm70k$
& $239\pm17$
& $50.0\%$
& $0$ \\

Qwen3 Max
& $5.0\pm0.0$
& $5/5$
& $3.2\pm1.6$
& $4907\pm623$
& $274\pm83$
& $52\text{k}\pm33\text{k}$
& $142\pm21$
& $29.4\%$
& $0$ \\
\bottomrule
\end{tabular}
}
\end{table*}

\begin{table}[t]
\centering
\caption{Main comparison results. }
\label{tab:main}
\setlength{\tabcolsep}{1.15pt}
\scriptsize
\begin{tabular}{@{}lccccc@{}}
\toprule
\textbf{Method}
& $\mathrm{Found}$
& $\mathrm{Compl.}$
& $\mathrm{Cycles}$
& \shortstack{$T_{\mathrm{comp}}$(s)}
& \shortstack{$E/\mathrm{vic}$(kJ)} \\
\midrule
{\texttt{Lawnmower}}
& $0.0\pm0.0$
& $0/5$
& $14.0\pm0.0$
& ---
& --- \\

{\texttt{Greedy-Coverage}}
& $0.0\pm0.0$
& $0/5$
& $32.0\pm0.0$
& ---
& --- \\

{\texttt{Bayesian Search}}
& $0.0\pm0.0$
& $0/5$
& $33.0\pm0.0$
& ---
& --- \\

{\texttt{Clue-Prior Sweep}}
& $2.6\pm0.9$
& $0/5$
& $14.0\pm0.0$
& ---
& $2186\pm1396$ \\

{\texttt{Clue-Prior Greedy}}
& $2.8\pm0.4$
& $0/5$
& $31.0\pm0.0$
& ---
& $1690\pm344$ \\

{\texttt{Clue-Prior Bayes}}
& $3.0\pm0.0$
& $0/5$
& $28.0\pm0.0$
& ---
& $1559\pm0$ \\

{\texttt{Ours}} (DS V4 Pro)
& $\mathbf{5.0\pm0.0}$
& $\mathbf{5/5}$
& $\mathbf{3.8\pm1.8}$
& $4661\pm457$
& $\mathbf{323\pm128}$ \\
\bottomrule
\end{tabular}
\end{table}

\subsubsection{Comparison Methods}
We compare the proposed solution with six non-LLM baselines and four LLM variants. 

\begin{itemize}[leftmargin=*,align=left]
\item \textbf{Baselines.} \rev{The six deterministic baselines span uninformed coverage, Bayesian filtering, and clue-derived spatial priors. The clue prior is constructed once from the last sighting, victim categories, elapsed time, terrain, water-seeking behavior, and the shared destination. Simulated trajectories and subgroup destinations are excluded.}
\begin{itemize}[leftmargin=*, nosep]
\small
    \item {\small\texttt{Lawnmower}}~\cite{choset2000coverage} covers the grid by sweeping each row sequentially from south to north.
    \item \rev{{\small\texttt{Greedy-Coverage}}~\cite{yamauchi1998frontier} dispatches each idle drone to its nearest feasible least-searched cell.}
    \item \rev{{\small\texttt{Bayesian Search}}~\cite{perez2018ant} starts from a uniform belief, applies terrain-aware motion prediction and negative-observation updates, and greedily selects cells by expected detection reward per unit travel-and-search time.}
    \item \rev{{\small\texttt{Clue-Prior Sweep}} orders the boustrophedon row strips once using their mass under the fixed clue prior.}
    \item \rev{{\small\texttt{Clue-Prior Greedy}} uses the same fixed prior throughout the mission and greedily trades off its probability mass against drone travel distance.}
    \item \rev{{\small\texttt{Clue-Prior Bayes}} initializes the Bayesian filter with subgroup-specific clue priors and subsequently applies the same motion and observation updates as {\small\texttt{Bayesian Search}}.}
\end{itemize}
\item \textbf{Variants.} \rev{Four LLM variants evaluate four design choices: two-stage decomposition, knowledge retrieval, explicit mobility reasoning, and drone-to-cell optimization.}
\begin{itemize}[leftmargin=*, nosep]
\small
    \item {\small\texttt{LLM-Direct}} directly asks the LLM to assign the idle drones to search cells. If both the initial response and one retry are invalid, the drones are assigned to their nearest feasible cells for that search cycle.
    \item {\small\texttt{w/o Knowledge Retrieval}} removes the retrieved domain knowledge $\mathcal{R}$ from the system message.
    \item {\small\texttt{w/o Mobility Reasoning}} removes Step~1; the LLM produces $\mathbf{w}(t)$ without reasoning about victim mobility.
    \item {\small\texttt{w/o Bottleneck}} reuses the per-cycle priority maps generated by {\small\texttt{Ours}} and replaces the bottleneck drone-to-cell matching with a uniformly random one-to-one assignment over the same selected cells.
\end{itemize}
\end{itemize}

\subsubsection{Evaluation Metrics} \rev{We evaluate the comparison methods, ablations, and LLM backbones based on five random seeds (42--46) with each run capped at 40 search cycles. Particularly, we report three groups of metrics.} % to evaluate mission effectiveness, resource consumption, and LLM output quality.
\begin{itemize}[leftmargin=*, nosep]
\small
    \item \rev{\textit{Mission effectiveness metrics} include i) victims found ($\mathrm{Found}$), ii) seeds achieving full completion ($\mathrm{Compl.}$), and iii) completion time $T_{\mathrm{comp}}$. The mean and standard deviation of $T_{\mathrm{comp}}$ are computed over completed seeds only; ``---'' means that no seed completes the mission.}
    \item \rev{\textit{Resource cost metrics}, reported as the mean and standard deviation across the five seeds, include i) search cycles ($\mathrm{Cycles}$), ii) energy per found victim ($E/\mathrm{vic}$), iii) total tokens per run ($\mathrm{Tok/run}$), and iv) average end-to-end agent computation time per search cycle ($T_{\mathrm{agent}}/\mathrm{cyc.}$).
    % For every method, the latter is first computed for each seed by dividing its total agent computation time by its number of cycles and then averaged over the five seeds.
    }
    \item \rev{\textit{Output quality metrics} include i) the illegal-output rate ($\mathrm{Illegal}$), computed as the fraction of all LLM responses across the five seeds that place priorities or assignments on impassable cells; and ii) the total number of search cycles ($\mathrm{Fb.}$) across the five seeds that invoke the fallback.}
\end{itemize}
%examine backbone and dispatch-batch sensitivity, and report mission effectiveness, resource cost, and output quality over five random seeds.

% A run terminates when all five victims are found, no feasible drone dispatch remains, or the cycle cap is reached.

\begin{figure*}[t]
\centering
\captionsetup{skip=0pt}
\includegraphics[width=\textwidth, trim=0 3mm 0 0,clip]{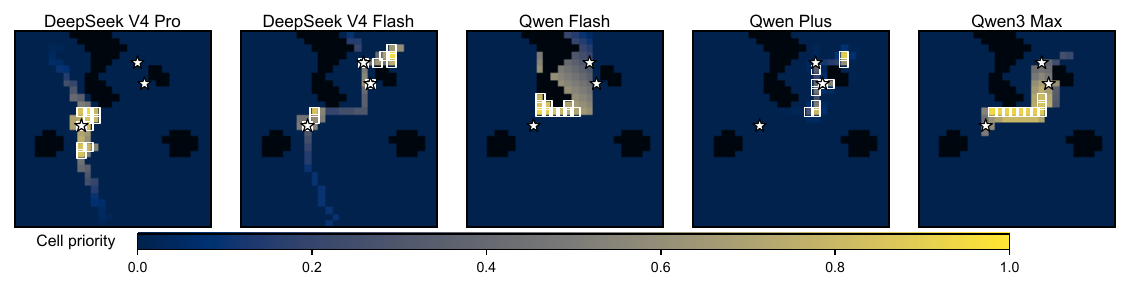}
\caption{\footnotesize Initial priority maps generated by five LLM backbones for seed 42 
%at the first decision time. % ($t=3600$~s, 
before any search observation. %). 
Cell color represents the inferred priority, white squares mark the top-$B$ ($B=10$) high-priority reachable cells, and stars mark the simulated victim-subgroup locations (hidden from the agent).}
\label{fig:initial-priority}
\end{figure*}
\vspace{-0.1cm}

\subsection{Results}
\label{sec:main}

\vspace{-0.1cm}

% This section shows the comparison results among our solution, baselines, variants, and multiple backbones.
% % , 
% % where each method is evaluated over five random seeds, $T_{\mathrm{comp}}$ is summarized over completed seeds only, and $\mathrm{Found}$ reports the number of victims found out of five seeds
% Meanwhile, our solution is examined against the initial priority-map behavior, a representative search trajectory, and dispatch-batch sensitivity.

This section presents comparisons of our solution with baselines, ablated variants, and multiple LLM backbones, together with analyses of the initial priority maps, a representative search trajectory, and dispatch-batch sensitivity.

\subsubsection{Performance Comparison}

\begin{figure*}[t]
\vspace{-0.5cm}
\centering
\includegraphics[width=18cm, trim=0 1mm 0 0,clip]{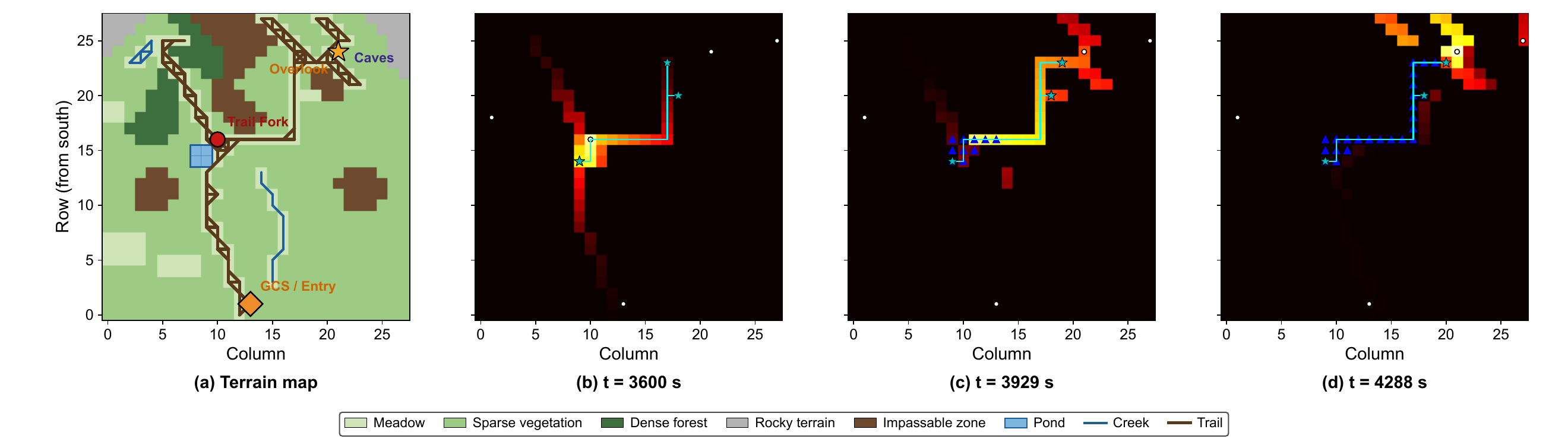}
\caption{\footnotesize \rev{Case study of the proposed agent solution. (a) shows the terrain map with four annotated landmarks. (b)--(d) show the inferred cell priorities at three reference moments. Heatmap intensity encodes $w_k$, cyan traces show victim paths, and blue triangles mark previously dispatched search cells.}}
\label{fig:case}
\end{figure*}

\cref{tab:main} compares our solution with baselines. Bold marks the best result in columns that permit direct comparison. The comparison highlights the benefit of reasoning over contextual clues and updating cell priorities across search cycles. The uninformed {\small\texttt{Lawnmower}}, {\small\texttt{Greedy-Coverage}}, and {\small\texttt{Bayesian Search}} baselines find no victim. Their inefficient assignments exhaust the available energy before the 40-cycle limit is reached. The three clue-prior baselines improve the average number of victims found to $2.6$--$3.0$, but none completes the mission. In contrast, {\small\texttt{Ours}} finds all five victims under all five seeds, requiring only $3.8$ search cycles and $323$~kJ per victim on average. These results show that a one-time clue-derived prior improves search focus, whereas updating the priorities with elapsed time and accumulated observations allows the search to adapt to victim movement and locate all victim subgroups.

\subsubsection{Ablation Study}
As shown in \cref{tab:ablation}, the two-stage policy turns LLM reasoning into more reliable and resource-efficient scheduling than direct assignment. 
{\small\texttt{LLM-Direct}} completes only two of the five runs. Compared with {\small\texttt{Ours}}, {\small\texttt{LLM-Direct}} nearly doubles the illegal-output rate from $26.3\%$ to $52.1\%$ and triggers six fallbacks, while {\small\texttt{Ours}} triggers none, showing that LLMs struggle to directly make scheduling decisions under constraints. {\small\texttt{w/o Knowledge Retrieval}} completes four of the five runs and increases the illegal-output rate from $26.3\%$ to $38.8\%$, indicating that task-specific knowledge helps the LLM reason about the likely locations of unfound victims. {\small\texttt{w/o Mobility Reasoning}} completes only one of the five runs and triggers five fallbacks, showing that directly generating cell priorities in a single step is unreliable. Compared with {\small\texttt{w/o Bottleneck}}, bottleneck matching reduces the mean worst-case outbound travel time per cycle by $16.1\%$, demonstrating that the drone-to-cell optimizer primarily shortens the slowest dispatch in each cycle.

% {\small\texttt{Ours}} also avoids final fallback in every cycle because feasibility filtering retains enough valid priorities for the optimizer. The ablations further distinguish the roles of the two reasoning components. Removing knowledge retrieval reduces completion from five to four seeds and increases the mean cycle count from $3.8$ to $9.6$, whereas removing mobility reasoning reduces completion to one seed and raises energy to $1854$~kJ per victim. 
% These results confirm that both components contribute to performance, with explicit mobility reasoning having the larger effect on mission completion.

% \vspace{-0.1cm}

\subsubsection{Cross-Backbone Analysis}
%\rev{We separately evaluate whether the proposed two-stage architecture transfers across LLM backbones and qualitatively compare each backbone's priority map with the locations of still-unfound victim subgroups.}

\cref{tab:backbone} demonstrates that the two-stage policy transfers across multiple backbones, although mission reliability and resource cost remain backbone-dependent. DeepSeek V4 Pro, Qwen3 Max and Qwen Plus complete all five seeds, while DeepSeek V4 Flash completes four and Qwen Flash completes none. DeepSeek V4 Flash delivers less reliable mission performance and incurs higher resource costs than DeepSeek V4 Pro. Qwen3 Max and Qwen Plus require $3.2$ and $5.6$ cycles on average, respectively, close to the $3.8$ cycles required by DeepSeek V4 Pro. However, their per-cycle agent computation times increase from $22$~s to $142$~s and $239$~s, resulting in longer mission completion times of $4907$~s and $6426$~s. Importantly, Qwen Plus reaches an illegal-output rate of $50.0\%$, comparable to {\small\texttt{LLM-Direct}}, but the optimizer in Stage~2 filters invalid priorities and enforces scheduling constraints before execution. Qwen Flash performs worst, finding only $1.6\pm0.9$ victims and completing no seed. With a $93.0\%$ illegal-output rate, all runs reach the 40-cycle limit, consuming $566$k tokens per run and $2201$~kJ per found victim despite only $19$~s of agent computation per cycle.

% Between the two fully completing backbones, DeepSeek V4 Pro reduces energy from $470$ to $323$~kJ per victim, tokens from $114$k to $32$k per run, and agent computation from $239$ to $22$~s per cycle, supporting its use as the default backbone. 
% The comparison also shows that illegal rate alone does not determine completion: Qwen Plus completes all five seeds despite a $50.0\%$ illegal rate, whereas DeepSeek V4 Flash completes only four at $12.5\%$. Mission performance therefore depends on the spatial content of the feasible priorities retained after filtering and how those priorities evolve over successive cycles.
% \IfFileExists{images/exp4_backbone_alignment.pdf}{
% \rev{For the spatial comparison, all five maps use the seed-42 initial decision at $t=3600$~s, so every backbone receives identical context and search history.}
%\Needspace{10\baselineskip}

\cref{fig:initial-priority} shows that the same contextual input produces different spatial distributions of the initial priorities across backbones. DeepSeek V4 Flash includes all three victim-subgroup locations among its top-$B$ cells, while DeepSeek V4 Pro and Qwen Plus each include one. Neither Qwen Flash nor Qwen3 Max directly covers a victim-subgroup location, but Qwen3 Max extends its top-$B$ cells along the trail toward the northern victim region, whereas Qwen Flash confines them to a short central segment. This difference is consistent with their mission performance: Qwen3 Max completes all five seeds, while Qwen Flash completes none.

\vspace{-0.05cm}
\subsubsection{Case Study}

\rev{We next examine how the inferred cell priorities evolve with victim movement and accumulated search observations during a completed mission.}
\rev{\cref{fig:case} visualizes how the inferred priorities remain aligned with the locations and movement routes of the still-unfound victims during the seed-42 run. At $t=3600$~s, the highest-priority region follows the trail toward the pond and Trail Fork, and includes the actual location of Family A. The first dispatch consequently finds both family members at $t=3907$~s. After these detections, the map at $t=3929$~s shifts eastward along the trail toward the campsite region occupied by Family B, which is subsequently found at $t=4280$~s. At $t=4288$~s, only Youth C remains unfound. The priority map then concentrates on cells around the victim's current position and the route toward the overlook; the following dispatch finds the final victim at $t=4488$~s. This progression reveals the mechanism behind the successful run: each time the high-priority region covers the current location of a remaining victim subgroup, the following dispatch discovers that subgroup, allowing all five victims to be found within four cycles.}
\vspace{-0.05cm}
\subsubsection{Dispatch-Batch Sensitivity}
\begin{figure}[t]
\centering
\includegraphics[width=\columnwidth]{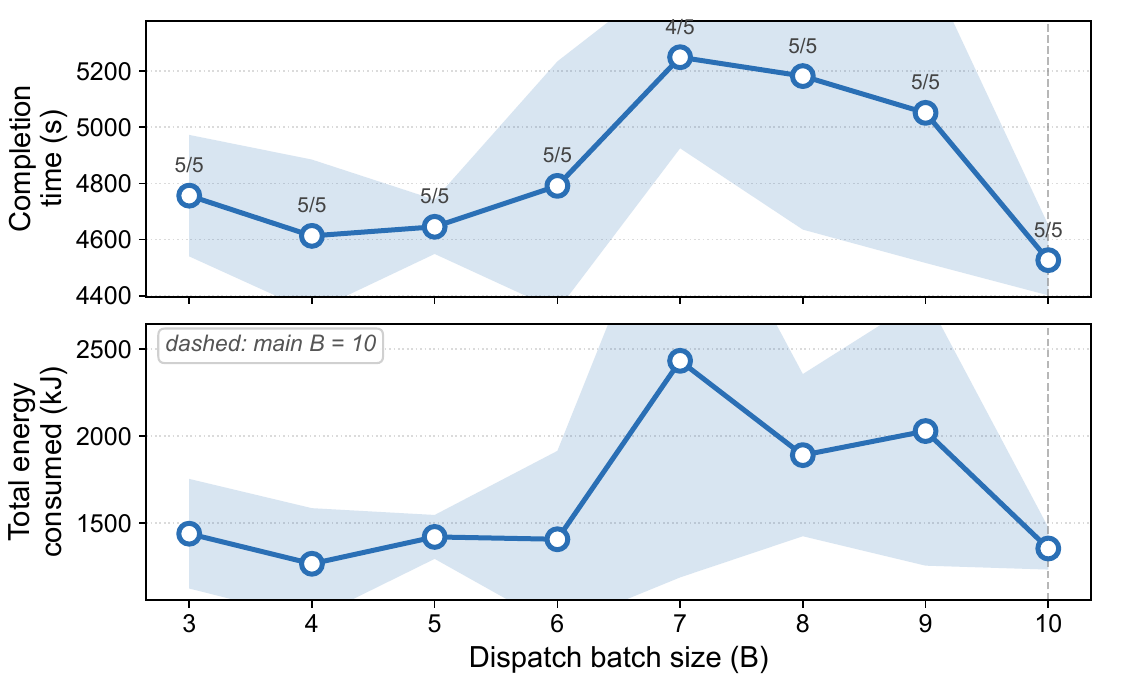}
\caption{\footnotesize \rev{Sensitivity to dispatch batch size $B$ with $N=10$. \textbf{Top:} completion time, with each label showing completed seeds. \textbf{Bottom:} total fleet energy. Curves show mean values and shaded bands show one standard deviation. The dashed line marks the default $B=10$.}}
\label{fig:bsweep}
\vspace{-0.3cm}
\end{figure}
We next evaluate how the number of drones dispatched per decision cycle affects mission completion and fleet energy. We fix $N=10$ and vary $B$ from $3$, the number of victim subgroups, to $10$, the fleet size. To ensure a comparable workload across batch sizes, we allow approximately 200 drone dispatches for each setting. For each batch size $B$, we therefore cap the number of search cycles at $\lceil 200/B\rceil$. Because this study uses different parameter settings, we rerun all batch-size configurations instead of reusing the results reported in the main comparison.
% We fix $N=10$ and a common budget of approximately 200 assignments. Specificly, we vary $B$ from $3$, the number of victim subgroups, to $10$, the fleet size. For each batch size $B$, we therefore cap the number of search cycles at $\lceil 200/B\rceil$. 
% The completion time uses completed seeds only. We fix $N=10$, use the common 200-assignment budget defined above, and keep the map, target-generation seeds, and LLM backbone unchanged across batch sizes. 
As shown in \cref{fig:bsweep}, seven of the eight settings complete all five seeds, while $B=7$ completes four and exhibits the largest energy variation. Among the fully completing settings, $B=10$ achieves the lowest mean completion time of $4527$~s, whereas $B=4$ minimizes mean total energy at $1266$~kJ. The settings $B=8$ and $B=9$ are dominated by $B=10$ in both completion time and energy consumption. These results demonstrate that batch size controls a genuine time--energy tradeoff: changing $B$ alters both the number of dispatched drones and the cells selected in each cycle, rather than providing a monotonic gain from additional parallelism.

%\Needspace{10\baselineskip}
%\subsection{Key Findings}
%\label{sec:discussion}

\textbf{Summary:} The results lead to three main findings. First, mission completion time is jointly determined by the number of search cycles and the duration of each cycle: more accurate cell priorities can reduce the required cycles by increasing the likelihood of searching the victims' actual locations, while, when different reasoning methods produce similar priorities and require similar numbers of cycles, assignment optimization can further reduce completion time by shortening the outbound travel time in each cycle. Second, unstructured contextual clues provide useful information for narrowing the search area, but a fixed clue-derived prior is insufficient when victims continue moving and new observations become available. Third, the main value of the LLM lies in its ability to combine contextual clues, domain knowledge, terrain constraints, elapsed time, and accumulated observations to continuously update search priorities; the optimizer then complements this reasoning by converting the priorities into energy-feasible and time-efficient drone assignments.
\vspace{-0.1cm}

\section{Conclusion}
\label{sec:conclusion}
\vspace{-0.1cm}
\rev{We developed an agent-aided drone scheduling system for victim search missions. The system performs efficient global drone scheduling using a joint LLM and optimization agent solution. The solution interprets unstructured victim clues and performs constrained drone scheduling in two sequential stages. First, the LLM reasons over a structured prompt to generate the priority vector of search cells. A drone-to-cell optimizer is then invoked to generate the scheduling policy while minimizing travel time. Simulations confirmed that interpreting contextual information is essential for mission completion, and the ablations validated the contributions of the individual design components.} 
Our future work will validate the proposed system in real-world victim-search missions using physical drones and geographic terrain data. Field experiments will further examine its reliability under practical factors not fully captured by the current simulations, such as sensing errors, communication delays, localization uncertainty, wind disturbances, and variable energy consumption.

\vspace{-0.1cm}

{\appendices
\section{}
\label{proof_the1}
\vspace{-0.1cm}
\textit{Derivation of the expected-cycle upper bound:}
Let $C_m$ denote the cycle in which victim $m$ is first confirmed, so that the total number of search cycles is $C=\max_{1\leq m\leq M}C_m$. Let $k_m(t_c)$ denote the cell occupied by victim $m$ at the actual execution time of the corresponding search task in cycle $c$, and let $\operatorname{Top}_B(\mathbf w(t_c);\mathcal R^*(t_c))$ denote the $B$ cells with the highest priorities within the reachable candidate pool $\mathcal R^*(t_c)$. For an unfound victim $m$, we define the quality (i.e., the probability's correctness) in the priority vector as $\rho_{m,c}\!\left(\mathbf w(t_c)\right)=\Pr\!\left(
k_m(t_c)\in
\operatorname{Top}_B\!\left(
\mathbf w(t_c);
\mathcal R^*(t_c)
\right)
\mid
C_m\ge c,\mathsf O(t_c),\mathsf B
\right).$
%\begin{equation}
% \Pr\!\left(
% k_m(t_c)\in
% \operatorname{Top}_B\!\left(
% \mathbf w(t_c);
% \mathcal R^*(t_c)
% \right)
% \mid
% C_m\ge c,\mathsf O(t_c),\mathsf B
% \right).
% \label{eq:rho_priority}
% \end{equation}
A larger $\rho_{m,c}(\mathbf w(t_c))$ means that the cells ultimately selected for search are more likely to include the actual cell of victim $m$.
% To isolate the performance of cell prioritization, we consider the case in which these $B$ cells can all be searched in the corresponding cycle.
Suppose that, for every reachable system state, every cycle $c$, and every victim $m$ with $C_m\ge c$, $\rho_{m,c}(\mathbf w(t_c))\ge\underline{\rho}>0$, while the detector satisfies $p_{\mathrm d}(\mathrm{terr})\ge\underline p_{\mathrm d}>0$ for every traversable terrain type. Then, the probability of an unfound victim being confirmed in each cycle is at least $\underline{\rho}\underline p_{\mathrm d}$, giving $\Pr_{\pi}(C_m>c)\leq(1-\underline{\rho}\underline p_{\mathrm d})^c$. Since $C>c$ implies that at least one victim remains unconfirmed, the union bound gives $\Pr_{\pi}(C>c)\leq M(1-\underline{\rho}\underline p_{\mathrm d})^c$. Since $0<\underline{\rho}\underline p_{\mathrm d}<1$, applying the tail-sum identity $\mathbb E_{\pi}[C]=\sum_{c=0}^{\infty}\Pr_{\pi}(C>c)$, we have 
\vspace{-0.35cm}

{\small
\begin{equation}
\mathbb{E}_{\pi}\!\left[\max_{1\leq m\leq M}C_m\right]
\leq
\left\lceil
\frac{\ln M}{-\ln\!\left(1-\underline{\rho}\underline p_{\mathrm d}\right)}
\right\rceil
+
\frac{1}{\underline{\rho}\underline p_{\mathrm d}}.
\label{eq:completion_cycle_bound}
\end{equation}}
\vspace{-0.35cm}

\noindent Thus, an upper bound for the expected completion-cycle is proved.
\hfill $\blacksquare$
}

\vspace{-0.5cm}
\bibliographystyle{IEEEtran}
\vspace{-0.1cm}
\bibliography{ref}

\end{document}